\documentclass[12pt,dvips]{article}
\usepackage{amsmath,graphics,epic,eepic,flafter,epsfig}

\catcode`\@=11

\@addtoreset{equation}{section}
\def\theequation{\arabic{section}.\arabic{equation}}

\catcode`\@=11
\def\thesection{\arabic{section}.}

\def\appendix{\setcounter{section}{0}
        \def\thesection{Appendix.}
        \def\theequation{\Alph{section}.\arabic{equation}}}
\def\section{\@startsection{section}{1}{\z@}{3.5ex plus 1ex minus
   .2ex}{2.3ex plus .2ex}{\large\bf}}

\newcommand{\captionfonts}{\small}
\makeatletter  
\long\def\@makecaption#1#2{%
  \vskip\abovecaptionskip
  \sbox\@tempboxa{{\captionfonts #1: #2}}%
  \ifdim \wd\@tempboxa >\hsize
    {\captionfonts #1: #2\par}
  \else
    \hbox to\hsize{\hfil\box\@tempboxa\hfil}%
  \fi
  \vskip\belowcaptionskip}
\makeatother   

\renewcommand{\Vec}[1]{\vec{\mathbf #1}}
\newcommand{\etahat}{\ensuremath{\boldsymbol{\h{\boldsymbol{\eta}}}}}
\newcommand{\psihat}{\ensuremath{\boldsymbol{\h{\boldsymbol{\psi}}}}}
\newcommand{\yhat}{\ensuremath{\boldsymbol{\h{y}}}}
\newcommand{\rhohat}{\ensuremath{\boldsymbol{\h{\boldsymbol{\rho}}}}}
\newcommand{\phihat}{\ensuremath{\boldsymbol{\h{\boldsymbol{\phi}}}}}
\newcommand{\divergence}[1]{\del\cdot\Vec{#1}}
\newcommand{\Ep}{\ensuremath{\eta^{\scriptscriptstyle+}}}
\newcommand{\Em}{\ensuremath{\eta^{\scriptscriptstyle-}}}
\newcommand{\Ap}{\ensuremath{a^{\scriptscriptstyle+}}}
\newcommand{\Am}{\ensuremath{a^{\scriptscriptstyle-}}}
\newcommand{\sss}{\scriptscriptstyle}

\newcommand{\Pgt}{\ensuremath{\Phi_{\scriptscriptstyle>}}}
\newcommand{\Plt}{\ensuremath{\Phi_{\scriptscriptstyle<}}}
\newcommand{\Jgt}{\ensuremath{\Vec{J}_{\scriptscriptstyle>}}}
\newcommand{\Jlt}{\ensuremath{\Vec{J}_{\scriptscriptstyle<}}}

\newcommand{\h}[1]{\ensuremath{{\bf\hat{#1}}}}        
\def\fpar#1#2{\frac{\partial#1}{\partial#2}}          
\newcommand{\Nabla}{\boldsymbol{\nabla}}
\newcommand{\del}{{\Vec{\Nabla}}}     
\newcommand{\curl}[1]{\ensuremath{\del\times\Vec{#1}}}  

\begin{document}

\begin{titlepage}

\ifx\mode\ugly
	\vspace{.5in}
\fi
\begin{flushright}
30 June 2000\\
\end{flushright}
\vspace{.5in}

\begin{center}
{\large\bf The Effect of Electric Fields In A Classic Introductory Physics Treatment of Eddy Current Forces}

\vspace{.3in}
{P.\  J.\  Salzman}\\
{\small\it Department of Physics}\\
{\small\it University of California}\\
{\small\it Davis, California 95616 USA}\\
{\small\it psalzman@dirac.org}\\[8pt]
{John\  Robert  Burke}\\
{\small\it Department of Physics and Astronomy}\\
{\small\it San Francisco State University}\\
{\small\it San Francisco, California 94132 USA}\\
{\small\it burke@stars.sfsu.edu}\\[8pt]
{Susan\ M.\ Lea}\\
{\small\it Department of Physics and Astronomy}\\
{\small\it San Francisco State University}\\
{\small\it San Francisco, California 94132 USA}\\
{\small\it lea@stars.sfsu.edu}\\
\end{center}

\vspace{.5in}
\begin{center}
{\bf Abstract}\\
\end{center}\penalty4000\begin{center}\penalty4000
\begin{minipage}{5in}
{\footnotesize

A simple model of eddy currents in which current is computed solely from magnetic forces acting on electrons proves accessible to introductory students and gives a good qualitative account of eddy current forces.  However, this model cannot be complete; it ignores the electric fields that drive current outside regions of significant magnetic field.  In this paper we show how to extend the model to obtain a boundary value problem for current density.  Solution of this problem in polar coordinates shows that the electric field significantly affects the quantitative results and presents an exercise suitable for upper division students.  We apply elliptic cylindrical coordinates to generalize the result and offer an exercise useful for teaching graduate students how to use non-standard coordinate systems.  

}
\end{minipage}
\end{center}
\end{titlepage}


\section{Introduction}
Every student of Electricity and Magnetism learns that Lenz's Law predicts a
force that opposes the motion of a conductor passing through a non-uniform
magnetic field.  Motion of the conductor's  free charge through the field
results in magnetic forces that drive current in the conductor.   This current,
in turn,  interacts with the $\Vec{B}$ field and results in a net magnetic force
acting on the conductor. The current is called an {\it eddy current\/}.

A classic classroom demonstration of eddy currents is a swinging metallic
pendulum that passes through the field of a strong magnet.   Eddy currents
within the conductor damp the oscillation rapidly.   When the conductor is 
replaced by another with holes, the eddy currents are impeded from circulating 
and the damping effect becomes very small.  The currents cease in this case 
because a Hall electric field develops that balances the magnetic force acting 
on the free charge.

We can estimate the eddy-current force acting on the conductor by using a few
simplifying assumptions\footnote{Susan M. Lea, John R. Burke \textit{Physics: 
The Nature of Things} (Brooks/Cole, 1997), p. 974.}$^{,}$\footnote{R. K. 
Wangsness, \textit{Electromagnetic Fields}, (Wiley, New York, 1986) 2nd ed.  
See prob 17--14 p. 282.}$^{,}$\footnote{For a careful discussion of the much 
more sophisticated theory due to Maxwell and an extensive list of references 
see W. M. Saslow, ``Maxwell's theory of eddy currents in thin conducting sheets,
and applications to electromagnetic shielding and MAGLEV'' Am. J. Phys.
\textbf{60}, 693--711 (1992).}.  First, model the conductor as a very large 
plane sheet passing between circular magnet poles of radius $a$.  Then, 
idealize the magnetic field as uniform in the cylindrical volume between the 
magnet poles and dropping abruptly to zero outside that volume.  Figure 1 
illustrates the model in a view perpendicular to the conducting sheet.

\vspace{.5cm}
\begin{centering}
\begin{figure}[!ht]
\centering
\hskip -5cm
\begin{picture}(300,150)(-215,-75)
\drawline(-150,-75)(150,-75)(150,75)(-150,75)(-150,-75)
\put(0,0){\circle{75}}
\put(0,0){\vector(0,1){55}}               
\put(3,48){\small \h{y}}                  
\put(0,0){\vector(1,0){55}}               
\put(48,3){\small \h{x}}                  
\put(150,0){\vector(1,0){40}}             
\put(160,4){\small $v_{0}\h{x}$}          
\put(-30,-12){\tiny$\Vec{B}=-B_{0}\h{z}$}
\put(-8,-25){$\otimes$}
\put(-75,-50){\small $\Vec{B}=0$}
\put(0,0){\vector(1,1){26}}
\put(12,20){\small $a$}
\end{picture}

\begin{minipage}{5.5in}
\caption{\footnotesize Magnetic field penetrating a circular portion of a moving
conductor.}
\end{minipage}
\end{figure}
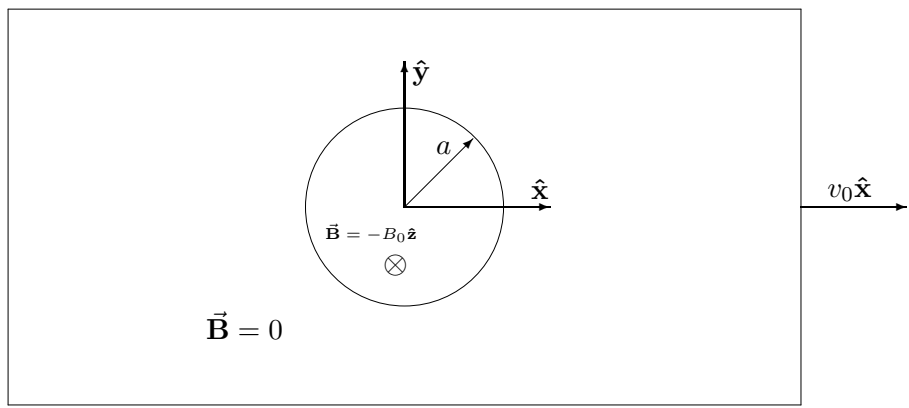

\end{centering}
\vspace{.5cm}

In this view, the magnetic field falls on a circular region of the conductor, is
uniform within the circle, and is zero outside the circle.  The conductor moves 
in the $\h{x}$ direction with speed $v_{0}$ and the magnetic field is given by 
$\Vec{B}=-B_{0}\h{z}$ within the circle of radius $a$.  By Ohm's Law, the current 
density is proportional to the force that drives it:
\begin{equation*}
   \Vec{J} = \sigma\Vec{v}\times\Vec{B} = \sigma vB \, \h{z},   \tag{1a}
	\label{one_ohm}
\end{equation*}
where $\sigma$ is the conductivity of the metal conductor.  Then the force
acting on a volume element $d\tau$ of the conductor within the field is:
\begin{equation*}
   d\Vec{F} = \Vec{J}\times\Vec{B}\, d\tau.
\end{equation*}
The net force acting on the conductor is
\begin{equation}
   \Vec{F} = -\sigma v_{0} B_{0}^{2} V\h{x}, \label{force} \tag{1b}
\end{equation}
\addtocounter{equation}{1}
where $V$ is the volume of the conductor exposed to the field.

This calculation correctly illustrates Lenz's Law and the dependence of the
force on velocity and magnetic field strength.  So, it gives a useful 
back-of-the-envelope estimate for the eddy current force.  However, it is a
somewhat naive estimate.  Once the current leaves the vicinity of the $\Vec{B}$
field, the model does not explain what causes the flow of free charge.  It
lacks an account of the Hall electric field (arising from charge distribution
on the surface of discontinuity of $\Vec{B}$) which drives the current outside
the $\Vec{B}$-field region, thus completing the current loops.  This $\Vec{E}$
field also opposes the current flow within the $\Vec{B}$-field region,
indicating that equation (1b) overestimates the force.  In this paper we
develop a method to account for this effect, and so to improve the estimate.

\section{The Exact Circle Problem}
Calculating the charge densities that give rise to electric fields driving
current in conductors is notoriously difficult\footnote{J.D. Jackson
``Surface charges on circuit wires and resistors play three roles'', Am. J.
Phys. \textbf{64}, 855--870 (1996).}, but is usually not necessary.  Here we 
can develop the calculation of current density as a two dimensional 
boundary-value problem using polar coordinates in the rest frame of the 
magnets.  We retain the simple model of the magnetic field from the 
introduction and, for now, model the plate as infinite in the
dimensions perpendicular to the field.  We also assume that the plate's
speed is sufficiently small that we can model the current distribution as a
quasi-steady state in the magnet frame.  The resulting problem
is a challenging but accessible  problem for upper division E\&M students.
In classic form, we observe that the current density is derivable from a
potential that satisfies Laplace's equation except at the magnetic field
boundary, develop the appropriate boundary conditions and solve via
expansion in eigenfunctions.

Then, the current density throughout is determined by Ohm's Law:
\begin{equation*}
   \Vec{J} = \sigma \left( \Vec{E} + \Vec{v}\times\Vec{B} \right).
\end{equation*}
Taking the curl of both sides and using a vector identity for $\del\times (
\Vec{v}\times\Vec{B} )$ yields:
\begin{equation}
   \Vec{\Nabla}\times\Vec{J} = -\sigma (\Vec{v}\cdot\Vec{\Nabla})\Vec{B}
\label{curlJ1}.
\end{equation}
Here we used the fact that the conductor's velocity and $\Vec{B}$ are constant
vectors in the magnet frame.  Inserting the assumed form for $\Vec{B}$ and
using polar coordinates for $\Vec{\Nabla}$, equation (\ref{curlJ1}) can be
rewritten as:
\begin{equation}
\Vec{\Nabla}\times\Vec{J} = -\sigma v_{0}B_{0}\cos(\phi)\delta(\rho-a)\h{z},
\end{equation}
where $\rho$ is the radial coordinate with origin at the center of the
magnetic field region.  Next, since no charge buildup is expected with
time, the
equation of continuity demands that $\Vec{J}$ be divergence free:
\begin{equation}
   \divergence{J} = 0.
\end{equation}
Since the current density is curl free except at $\rho=a$, it is the
gradient of some scalar potential on each side of $a$: $\Jgt = - \del\Pgt$ 
and $\Jlt = - \del\Plt$, where the subscripts $>$ and $<$ refer respectively 
to regions outside and inside the boundary.  Thus $\del^2 \Phi=0$ everywhere 
except at the magnetic field boundary, and we may proceed with standard 
methods for solving Laplace's equation.

The boundary conditions on the components of $\Vec{J}$ perpendicular and
parallel to the boundary follow from the divergence and curl of current
density.  This is a standard calculation, with the results
\begin{gather*} 
J_{\sss >,\perp}(\Ap) - J_{\sss <,\perp}(\Am) = 0; \\ 
J_{\sss <,\parallel}(\Ap) - J_{\sss <,\parallel}(\Am) = -\sigma v_0 B_0 \cos(\phi).
\end{gather*}
In terms of potential, the boundary conditions are:
\begin{gather} 
   \partial_{\rho}\Pgt(\rho=\Ap) - \partial_{\rho}\Plt(\rho=\Am) = 0; \tag{5a} \label{the_a} \\ 
   \partial_{\phi}\Pgt(\rho=\Ap) - \partial_{\phi}\Plt(\rho=\Am) = \sigma a v_0
	B_0 \cos(\phi) \tag{5b} \label{the_b}.
\end{gather}
\addtocounter{equation}{1}
The problem is now completely specified, and we proceed by expanding the
potential in eigenfunctions of the Laplace operator in polar coordinates.
\begin{equation} 
\Phi(\Vec{x}) = \begin{cases}
   \Plt = \sum_{n=0}^{\infty}\rho^n    \left[ A_n\sin(n\phi) + B_n\cos(n\phi) 
		\right] & \rho < a \\
   \Pgt = \sum_{n=0}^{\infty}\rho^{-n} \left[ A_n\sin(n\phi) + B_n\cos(n\phi)
		\right] & \rho > a.
\end{cases} \label{expansion}
\end{equation}
From the boundary condition for $J_{\sss\parallel}$ (eqn \ref{the_b}) and the
orthogonality of the trigonometric functions, we see that the $n=1$ terms are
the only non-zero terms in the sums.  Furthermore, only the $A$ coefficients
are non-zero.  The boundary conditions on the potential now give two equations
for the coefficients.  We find:
\begin{equation*}
\Phi(\Vec{x}) = \begin{cases}
\Plt = -\frac{1}{2} \sigma v_0 B_0 \rho\sin(\phi) & \rho < a \\
\Pgt = +\frac{1}{2} \sigma v_0 B_0 \frac{a^2}{\rho}\sin(\phi) & \rho > a.
\end{cases}
\end{equation*}
We can take gradients to obtain the current density:
\begin{equation}
\Vec{J}(\Vec{x}) = \begin{cases}
\Jlt = -\del\Plt = \frac{1}{2}\sigma v_0 B_0 \,\yhat & \rho < a \\
\Jgt = -\del\Pgt = \frac{1}{2}\sigma v_0 B_0 \left( \frac{a}{\rho} \right)^{2}
		\left[ \sin(\phi)\rhohat - \cos(\phi)\phihat  \right] & \rho > a.
\end{cases}
\label{InfiniteCircleCurrent}
\end{equation}
We still find a uniform current density within the magnetic field region.
The current outside the field region follows a classic dipole pattern.
The corresponding electric field that drives current in the region $\rho>a$
and opposes it in the region $\rho<a$ is found from
\begin{align}
\Vec{E} &= \frac{\Vec{J}}{\sigma} - \Vec{v} \times \Vec{B} \\
	    &= \begin{cases}
	-\frac{1}{2} v_0 B_0 \, \yhat    & \rho < a \\[5pt]
	\frac{1}{2} v_0 B_0 \left(\frac{a^2}{\rho}\right) \left[
		\sin(\phi)\,\rhohat - \cos(\phi)\,\phihat \right]   & \rho > a
	\end{cases}
\end{align}
The charge density $\sigma_c$ that gives rise to this field is localized at
$\rho=a$ and is found from the standard boundary condition:
\begin{align}
\sigma_c
&= \varepsilon_0 \, \rhohat \cdot \left[ \Vec{E}_> - \Vec{E}_< \right] \\
&= \varepsilon_0 v_0 B_0 \sin(\phi)
\end{align}
For a field $B_0 = 1$T and a plate speed of $v_0=1\,$m/s, the charge density is
of order $\sigma_c \sim 10^{-11}$ C/m$^2$.

Comparison equation (\ref{InfiniteCircleCurrent}) with equation (\ref{one_ohm})
shows that the current, and hence the net force acting on the conductor is half
that predicted by the naive model.  Such a simple result, in contrast with the
complex correction one might have expected, rasies the issue whether a
correction factor of $1/2$ is generally correct or specific to the circuclar
field geometry.  We investigate that question in the following sections.


\section{Elliptical Magnetic Field Region}
The result for a circular magnetic field geometry demonstrates that electric 
field has a significant effect on eddy current flow.  We were intrigued whether 
the factor of 1/2 reduction is a general result or special to the case of 
circular geometry.  To investigate this question, we solved the problem of an 
elliptically shaped magnetic field region with eccentricity $\epsilon$, as
shown in Figure~2.  The 
method follows the same outline as the circle problem except that we expand the 
potential in elliptic cylindrical coordinates, defined\footnote{Morse and
Feshbach, \textit{Methods of Theoretical Physics} (McGraw-Hill, New York,
1953), Vol. 1, p. 514.}$^{,}$\footnote{Moon and D.E. Spencer, \textit{Field 
Theory Handbook}, (Springer Verlag, Berlin, 1961), pp. 17--19.} in terms of 
Cartesian by:
\begin{equation}
	x= h\cosh(\eta)\cos(\psi)  \qquad  y= h\sinh(\eta)\sin(\psi)  \qquad  z= z.
\end{equation}
The unit vectors are given by
\begin{equation}
\h{x} = \frac{\etahat \sinh(\eta)\cos(\psi) - \psihat \cosh(\eta)\sin(\psi)}
	{\sqrt{\cosh^2(\eta) - \cos^2(\psi)}} \qquad
\h{y} = \frac{\etahat \cosh(\eta)\sin(\psi) + \psihat\sinh(\eta)\cos(\psi)}
  {\sqrt{\cosh^2(\eta) - \cos^2(\psi)}}.
\label{unitvector}
\end{equation}
The constant $h$ is the product of the semi-major axis $a$ and eccentricity
$\epsilon$ of the elliptical magnetic field region.  The boundary of the 
magnetic field is defined by the level curve
\begin{equation}
\eta = \eta_0 = \cosh^{-1}\left(\frac{1}{\epsilon}\right).  \label{etanaught}
\end{equation}
%
{
\setlength{\unitlength}{.8mm}
\begin{figure}[!ht]
\centering
\hskip -7cm
\begin{picture}(100,50)(-93,-25)
\drawline(50,25)(-50,25)(-50,-25)(50,-25)(50,25)
\put(0,0){\ellipse{60}{30}}
\put(0,0){\qbezier(50,24.5)(-8,0)(50,-24.5)}
\put(0,0){\qbezier(-50,24.5)(8,0)(-50,-24.5)}
\put(0,0){\dashline[-30]{3}(0,0)(44,22)}
\put(0,0){\dashline[-30]{3}(0,0)(44,0)}
\put(8,1.5){\makebox(0,0){$\scriptscriptstyle \phi=\pi/6$}}

\put (24, 0){\circle*{1}}
\put (24,-3){\makebox(0,0){$\scriptstyle q$}}
\put(-24, 0){\circle*{1}}
\put(-24,-3){\makebox(0,0){$\scriptstyle p$}}

\put(0,17){\makebox(0,0){$\scriptstyle \eta=\eta_0$}}
\put(40,17) {\rotatebox{30}{\makebox(0,0){$\scriptstyle \psi=\pi/6$}}}
\put(-40,17){\rotatebox{-30}{\makebox(0,0){$\scriptstyle \psi=5\pi/6$}}}
\put(-40,-17){\rotatebox{30}{\makebox(0,0){$\scriptstyle \psi=7\pi/6$}}}
\put(40,-17){\rotatebox{-30}{\makebox(0,0){$\scriptstyle \psi=11\pi/6$}}}

\put(12,-17){\makebox(0,0){$\scriptscriptstyle h$}}
\put(0,-18){\vector(1,0){24}}
\put(12,-21){\makebox(0,0){$\scriptscriptstyle a$}}
\put(0,-22){\vector(1,0){30}}

\end{picture}\break
\begin{minipage}{5.5in}
\caption{\footnotesize The magnetic field penetrates an elliptical portion of 
a moving
conductor.  The level curves of coordinate $\eta$ are ellipses with foci
$p$ and $q$, a distance $2h$ apart.  The level curves of $\psi$ are
semi-hyperbolae; the value of $\psi$ is the polar angle of the asymptote.
The boundary of the magnetic field is given by $\eta=\eta_{0}=\cosh^{-1}
(\frac{1}{\epsilon})$.  The semi-major axis of the boundary is $a$.}
\end{minipage}
\end{figure}
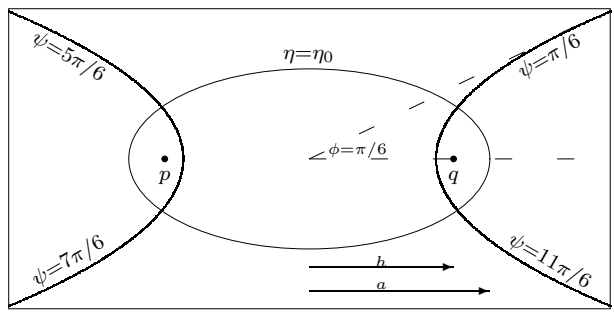
}

In these coordinates, the vector expressions for the divergence and 
curl of $\Vec{J}$ are unchanged:
\begin{equation*}
\divergence{J} = 0 \qquad \curl{J} = -\sigma (\Vec{v}\cdot\del)\,\Vec{B}(\eta)
\end{equation*}
where $\Vec{B}=-B_{0}\h{z}$ within the field region and zero outside.  Using 
equation (\ref{unitvector}) we may express $(\Vec{v}\cdot\del)\Vec{B}$ in this
coordinate system.  We find:
\begin{align*}
	\curl{J} 
	&=-\frac{\sigma v_0 \sinh(\eta)\cos(\psi)}{h[\cosh^{2}(\eta)-\cos^{2}(\psi)]}
		\fpar{\Vec{B}(\eta)}{\eta}, \\
	&=-\frac{\sigma v_0 B_0 \sinh(\eta)\cos(\psi)} {h[\cosh^{2}(\eta) - \cos^{2}
		(\psi)]} \delta(\eta-\eta_0) \, \h{z}.
\end{align*}
As before, conditions on the curl and divergence of $\Vec{J}$ lead to boundary 
conditions:
\begin{align*}
	J_{\sss >,\perp}(\eta_0) - J_{\sss <,\perp}(\eta_{0}) &= 0,  \\
	J_{\sss >,\parallel}(\eta_0) - J_{\sss <,\parallel}(\eta_0) &= 
		-\frac{\sigma v_0 B_0 \sinh(\eta_0)\cos(\psi)}{\sqrt{\cosh^{2}(\eta_0) 
			- \cos^2(\psi)}}.
\end{align*}
Once again we can make the argument that $\Vec{J}$ is the gradient of some
scalar potential $\Phi$ in the regions separated by $\eta=\eta_0$.  That is,
$\Jlt = -\del\Plt$ for $\eta < \eta_0$ and $\Jgt = -\del\Pgt$ for $\eta >
\eta_0$.  Components of current which lie perpendicular and parallel to the
curve bounding the $\Vec{B}$ region are then given by: 
\begin{align*}
	J_{\sss\perp}    &= -\del\Phi\cdot\etahat = -\frac{1}{h\sqrt{\cosh^2(\eta) - 
		\cos^2(\psi)}}\fpar{\Phi}{\eta}, \\
	J_{\sss\parallel}&= -\del\Phi\cdot\psihat = -\frac{1}{h\sqrt{\cosh^2(\eta) - 
		\cos^2(\psi)}}\fpar{\Phi}{\psi}.
\end{align*}
The boundary conditions on $\Vec{J}$ then give us the following boundary
conditions on the potential:
\begin{align*}
\partial_{\psi}\Pgt(\eta=\Ep_0) - \partial_{\psi}\Plt(\eta = \Em_0) &=
		\sigma h v_0 B_0 \sinh(\eta_0)\cos(\psi),\\[5pt]
\partial_{\eta}\Pgt(\eta=\Ep_0) - \partial_{\eta}\Plt(\eta = \Em_0) &= 0.
\end{align*}
The expansion of the potential in terms of eigenfunctions of the Laplace
operator in this coordinate system is\footnote{P. Moon and D.E. Spencer, 
\textit{ op. cit.}, 19.}:
\begin{equation}
\Phi(\Vec{x}) = \begin{cases}
	\Plt = \sum_{n=0}^{\infty} \left[ E_n \sinh(n\eta) + F_n \cosh(n\eta) \right]
		\left[ A_n \sin(n\psi) + B_n \cos(n\psi) \right] & \eta < \eta_0\\[5pt]
	\Pgt = \sum_{n=0}^{\infty} \left[ G_n e^{n\eta} + H_n e^{-n\eta} \right]
		\left[ C_n \sin(n\psi) + D_n \cos(n\psi) \right] & \eta > \eta_0.
\end{cases}
\label{FullBlownEllipticExpansion}
\end{equation}
Once again, orthogonality of the trigonometric functions ensures that only
$n=1$ terms will be non-zero. In the limit $\eta\rightarrow\infty$, we require
$\Pgt$ to remain finite, so we have $G_1 = 0$.  The limit $\eta\rightarrow0$
describes the portion of the x-axis between the two foci.  Here the curl and
divergence of $\Vec{J}$ are both zero, so that $\Vec{J}$ is continuous across
the x-axis.  Since $\eta$ increases away from the x-axis on both sides,
$\etahat$ changes direction discontinuously across the x-axis.  Thus,
continuity of $J_{\sss\perp}$ implies that $J_{\eta}\propto
\partial_{\eta}\Plt$ changes sign across the x-axis. This fact requires that
the $\cos(\psi)$ term in $\Plt$ be zero since only $\sin(\psi)$ is
discontinuous at $y=0$.  Now, $\psihat$ is also discontinuous across the x-axis
between the foci, so continuity of $J_{\sss\parallel}$ requires that
$\partial_{\psi}\Plt \propto \cos(\psi)$ either be discontinuous (which it
isn't) or be zero.  Thus $\partial_{\psi}\Plt=0$ as $\eta\rightarrow0$, which
requires that $F_1=0$.  Then, the boundary conditions at $\eta_0$ require
\begin{equation*}
\Phi(\Vec{x}) = \begin{cases}
	\Plt = -\frac{\sigma h v_0 B_0 \sinh(\eta_0)}{\cosh(\eta_0) + \sinh(\eta_0)}
		\sinh(\eta)\sin(\psi)  & \eta < \eta_0 \\
	\Pgt = \sigma h v_0 B_0 \sinh(\eta_0)\cosh(\eta_0) e^{-\eta}\sin(\psi) &
		\eta > \eta_0.
\end{cases}
\label{EllipticPotentialAfterBC}
\end{equation*}
We can take gradients to calculate the exact current density:
\begin{equation}
\Vec{J}(\Vec{x}) = \begin{cases} 
	\Jlt = \frac{\sigma v_0 B_{0} \sinh(\eta_0)}{\cosh(\eta_0) + \sinh(\eta_0)}
		\,\yhat & \eta < \eta_0 \\
	\Jgt = \sigma v_0 B_0 \sinh(\eta_0) \cosh(\eta_0) e^{-\eta} \left( 
		\frac{\etahat \sin(\psi) - \psihat \cos(\psi)}{\sqrt{\cosh^2(\eta) - 
		\cos^2(\psi)}} \right) & \eta > \eta_0.
\end{cases}
\end{equation}
Substituting for $\eta_0$ in terms of the $\Vec{B}$-field region's eccentricity 
(eqn \ref{etanaught}) we find:
\begin{equation}
\Vec{J}(\Vec{x}) = \begin{cases}
	\Jlt = \sigma v_0 B_0 \frac{\sqrt{1-\epsilon^2}}{1 + \sqrt{1-\epsilon^2}}
		\, \yhat & \eta < \cosh^{-1}(1/\epsilon) \\
	\Jgt = \sigma v_0 B_0 \frac{\sqrt{1-\epsilon^2}}{\epsilon^2} e^{-\eta} \left(
		\frac{\etahat \sin(\psi) - \psihat \cos(\psi)}{\sqrt{\cosh^2(\eta) - 
		\cos^2(\psi)}} \right) & \eta > \cosh^{-1}(1/\epsilon).
\end{cases}
\end{equation}
Again we find uniform current density in the magnetic field region.  The factor
of 1/2 reduction in current density found for the circular field turns out not
to be general.  It is replaced by the factor 
\begin{equation*}
	\frac{\sqrt{1-\epsilon^2}}{1+\sqrt{1-\epsilon^2}},
\end{equation*}
which, of course, has the limit $1/2$ as $\epsilon\rightarrow0$. The graph of
this function is shown as Figure~3.  Since the force acting on the sheet is
$\Vec{F}_{net} = - B_{0}|\Jlt|V \, \h{x}$ where $V$ is the volume exposed to
the magnetic field, the force also has the expected limit.  It is much more
intricate and much less crucial to establish that the expression for $\Jgt$
reduces to the circular results in the limit $\epsilon\rightarrow0$.  The
calculation is not given here, but a copy of it is available from the authors
upon request.


\section{The Effect Of Finite Conductor Size}

Once these two calculations are set up for infinite plates, it is easy to 
estimate the correction for finite plate size.  One changes the boundary 
condition from $|\Jgt|\rightarrow0$ at infinity to vanishing of the radial 
component of $\Jgt$ at a finite radial coordinate.  

For the circular case we take the plate to have a finite radius $R$.  A 
solution is only feasible for the time when the plate is centered on the 
magnetic field region, so the result offers only an order of magnitude 
estimate of the effects.

In eqn (\ref{expansion}) (the expansion of the potential) an extra term in 
$\Pgt$ that increases with $\rho$ is necessary to match the new boundary 
condition.  A straightforward calculation reveals:
\begin{equation}
	\Vec{J}(\Vec{x}) = \begin{cases}
		\frac{1}{2}\sigma v_0 B_0 \left(1-\frac{a^2}{R^2} \right) \yhat & 
		\rho < a \\[5 pt]
	\frac{1}{2} \sigma v_0 B_0 \left( \frac{a}{\rho R} \right)^2 \left[ (R^2 -
		\rho^2) \sin(\phi)\rhohat - (R^{2} + \rho^{2}) \cos(\phi) \phihat \right]
		& R > \rho > a.
\end{cases}
\end{equation}
One may quickly verify that these expressions have the correct limits (eqn
\ref{InfiniteCircleCurrent}) in the infinite conductor case ($R\rightarrow
\infty$).  The effect on the dipole current term is substantial near the
boundary.  Current in the magnetic field region is further reduced by the edge
effects, but by an insubstantial amount, unless the distance from the
center to the nearest edge is comparable to the radius of the field region.

A similar calculation is possible for elliptic cylindrical coordinates
with a border at $\eta=H$ (semi-major axis of the boundary is $A=a\epsilon\cosh
(H)$.  As in circular geometry, we augment the old conditions with the new 
condition that the elliptic-radial component of $\Jgt$ vanishes at the boundary:
$\partial_{\eta}\Phi(H,\psi)=0$.  We find:
\begin{equation*}
\Jlt(\Vec{x}) =  \sigma v_0 B_0 \frac{\sqrt{1-\epsilon^2}}{\epsilon^2} \left( 
	\sqrt{1 - \left( \frac{a\epsilon}{A} \right)^2 } - \sqrt{1-\epsilon^2} 
	\right) \qquad \eta < \cosh^{-1}(\frac{1}{\epsilon}).
\end{equation*}
Observing that $\sqrt{1-\epsilon^2}/(1+\sqrt{1-\epsilon^2}) = \sqrt{1 - 
\epsilon^2} (1-\sqrt{1-\epsilon^2}) / \epsilon^2$, we see that the effect on 
$\Jlt$ is to replace $1$ with $\sqrt{1-\left( \frac{a\epsilon}{A} \right)^2}$
in the last factor of the expression.  Again the correction is of order of the
square of the ratio  of magnet size to the plate dimension.

\section{Conclusion}

We have demonstrated that the electric field has a significant effect on the 
eddy-current force computed from a simple model.  The model gives the magnitude
of the force as
\begin{equation}
F = \sigma v_0 B_0^2 V f,
\end{equation}
where $V$ is the volume of conductor exposed to the magnetic field $B_0$,
$\sigma$ is the conductivity of the metal and $v_0$ is its speed relative
to the source of the magnetic field.  The factor $f$ is the correction due to
the electric field; $1/2$ for an infinite metal plate and circular magnet poles
and $\sqrt{1-\epsilon^2}/(1+\sqrt{1-\epsilon^2})$ for an infinite metal plate 
with elliptical magnetic poles.  
\begin{figure}[!ht]
\centering
\includegraphics{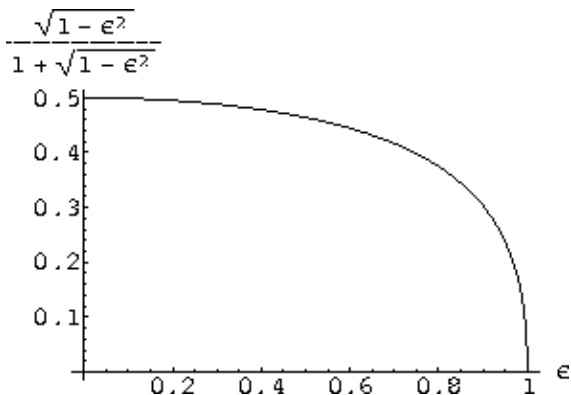}\break
\begin{minipage}{5.5in}
\caption{\footnotesize The reduction factor in the current density for an 
arbitrary ellipse as a function of the ellipse's eccentricity.}
\end{minipage}
\end{figure}
The first result follows from a boundary value problem accessible to an upper 
division student, while the second result requires boundary value techniques
that would be good training for a graduate 
student.  Corrections for finite plate size alter the result by terms of order
(magnet size / plate size)$^2$. 

A possible objection to this method is the need for assuming an abrupt
edge to the magnetic field region.  Burke and Lea have developed a method for
treating a more realistic model of the field\footnote{John R. Burke, Susan Lea
in preparation.}.  In the limit of zero separation of the magnet poles they
find $f=1/2$.  For a pole separation of one-tenth of the pole radius, they
find $f=0.39266$.

In all cases, $0\le f\le0.5$, though $0.5$ seems to be a robust approximation.

\section{Acknowledgements}

This work was supported in part by the Department of Energy under grant
DE-FG03-91ER40674.  It was also supported in part by The Portland Group,
Inc. (PGI) for the generous use of the PGI Workstation.

\end{document}